\documentclass[reprint,amsmath,twocolumn,amssymb,nobibnotes,aps,pra,superscriptaddress]{revtex4-1}

\usepackage[T1]{fontenc}
\usepackage{microtype}
\usepackage{graphicx}
\usepackage{dcolumn}
\usepackage{bm}
\usepackage{color}
\usepackage{color}
\usepackage{xspace}
\usepackage{braket}

\newcommand{\PT}{$\mathcal{PT}$\xspace}
\newcommand{\CPT}{$\mathcal{CPT}$\xspace}
\newcommand{\e}{\mathrm{e}}
\newcommand{\im}{\mathrm{i}}

\begin{document}

\title{Computing eigenfunctions and eigenvalues of boundary value problems
  with the orthogonal spectral renormalization method}

\author{Holger Cartarius}
\affiliation{Institut f\"ur Theoretische Physik 1, Universit\"at Stuttgart,
  Pfaffenwaldring 57, 70550 Stuttgart, Germany}
\author{Ziad H. Musslimani}
  \affiliation{Institut f\"ur Theoretische Physik 1, Universit\"at Stuttgart,
  Pfaffenwaldring 57, 70550 Stuttgart, Germany}
  \affiliation{Department of Mathematics, Florida State University, Tallahassee,
  FL 32306-4510}
\author{Lukas Schwarz}
\affiliation{Institut f\"ur Theoretische Physik 1, Universit\"at Stuttgart,
  Pfaffenwaldring 57, 70550 Stuttgart, Germany}
\author{G\"unter Wunner}
\affiliation{Institut f\"ur Theoretische Physik 1, Universit\"at Stuttgart,
  Pfaffenwaldring 57, 70550 Stuttgart, Germany}
\date{\today}

\begin{abstract} 
The spectral renormalization method was introduced in 2005 as an
effective way to compute ground states of nonlinear Schr\"odinger and
Gross-Pitaevskii type equations. In this paper, we introduce an
orthogonal spectral renormalization (OSR) method to compute ground and
excited states (and their respective eigenvalues) of linear and
nonlinear eigenvalue problems. The implementation
of the algorithm follows four simple steps: (i) reformulate the underlying
eigenvalue problem as a fixed point equation, (ii) introduce a renormalization
factor that controls the convergence properties of the iteration,
(iii) perform a Gram-Schmidt orthogonalization process in order to
prevent the iteration from converging to an unwanted mode; and (iv)
compute the solution sought using a fixed-point iteration. The
advantages of the OSR scheme over other known methods (such as
Newton's and self-consistency) are: (i) it allows the flexibility to
choose large varieties of initial guesses without diverging, (ii) easy
to implement especially at higher dimensions and (iii) it can easily handle
problems with complex and random potentials. The OSR method is
implemented on benchmark Hermitian linear and nonlinear eigenvalue 
problems as well as linear and nonlinear non-Hermitian \PT-symmetric models.
\end{abstract}

\maketitle

\section{Introduction}
In this paper, an orthogonal spectral renormalization (OSR) method is
proposed as a mean to compute ground and excited states for linear and
nonlinear boundary value problems, an application which is important for
quantum systems and beyond \cite{gilboa-image-processing,%
  JMathImagingVision.56.300}. The core idea is to recast the
eigenvalue problem as a fixed point equation which is then numerically
solved using a renormalized iterative scheme. The excited states are
computed using a Gram-Schmidt orthogonalization process whose sole
purpose is to avoid convergence to an undesired state. The proposed
algorithm is robust and easy to implement particularly on problems
where traditional methods (such as Newton's and self-consistency, see e.g.\
\cite{pang_computational_physics}) are either difficult to use or fail to
converge.  The advantages of the OSR scheme over other well established methods
are: (i) it allows the flexibility to choose large varieties of initial
guesses without diverging, (ii) easy to implement especially at higher
dimensions and (iii) it can easily handle problems with complex and random
potentials. The OSR method is implemented on typical {\it Hermitian} linear and
nonlinear eigenvalue problems. Examples include the linear harmonic and
anharmonic oscillators, one-dimensional particle in a box and the nonlinear
Schr\"odinger/Gross-Pitaevskii equation in the presence of a Hermitian
harmonic trap. In addition, the OSR scheme is used to compute the spectrum
of the \PT-symmetric Hamiltonian introduced originally by Bender and Boettcher
\cite{PhysRevLett.80.5243} and a BEC in a double-well potential with
gain and loss. The latter is modeled by the Gross-Pitaevskii equation in the
presence of a complex external potential.

The applicability of the method to non-Hermitian problems is of special
importance since in the last decade there has been an increased interest in
non-Hermitian quantum and optical systems, especially those that obey
the so-called \PT symmetry \cite{PhysRevLett.80.5243}. Generally
speaking, such systems can be described by a linear Schr\"odinger or
Gross-Pitaevskii type equation with the presence of an external
complex potential $V(x)$. Space-time reflection (\PT) symmetry implies
the relation $V(x)=\bar{V}(-x)$, where bar stands for complex
conjugation. The physical consequences of such symmetry have been
intensively studied in many branches of the physical
sciences. Extensive studies exist in theoretical physics, where they
cover fundamental questions in quantum mechanics
\cite{JMathPhys.40.2201,PhysLettA.264.108,JPhysA.43.145301,JPhysA.43.055307}
and new forms of quantum field theories \cite{PhysRevD.85.085001,%
  FortschrPhys.61.140}.  Recently, relations of \PT symmetry with topologically
nontrivial phases in many-body systems became important
\cite{PhysRevB.84.153101,PhysRevB.84.205128,JPhysCondensMatter.24.145302,%
  OptLett.11.1912,PhysRevLett.115.040402,PhysLettA.379.1213,%
  PhysRevA.89.062102,PhysRevA.92.012116,PhysRevA.95.053626,NatMater.16.433}.
Since a promising approach of realizing a genuine \PT-symmetric quantum system
are Bose-Einstein condensates, where atoms are removed from and added to the
condensed phase \cite{PhysRevLett.101.080402,FortschrPhys.61.124,%
  PhysRevA.90.042123,JPhysA.48.335302,PhysRevA.93.023624}, the study of \PT
symmetry became also important within the nonlinear Gross-Pitaevskii equation.
This has been done up to now in a broad variety of ways ranging from a two-mode
approximation \cite{JPhysA.41.255206,JPhysA.45.444015} to detailed
descriptions in position space \cite{FortschrPhys.61.124,JPhysA.48.335302,%
IntJTheorPhys.54.4054,JPhysA.41.244019,PhysRevA.95.053613}.

Since in many special cases a formal equivalence exists between the
Schr\"odinger equation and Maxwell's equations, the concept of \PT
symmetry can also be studied in electromagnetic waves such as
microwave cavities \cite{PhysRevLett.108.024101} or even in electronic
devices \cite{PhysRevA.84.040101}. The most dramatic advances have
been achieved in optics, where the notion of \PT symmetry can be used
to describe wave guides with complex refractive indices
\cite{PhysRevLett.101.080402,PhysRevLett.100.103904,PhysRevLett.100.030402,%
  OptLett.32.2632,PhysRevA.95.053868}, in which the first experimental
confirmation of \PT symmetry and \PT symmetry breaking were made possible
\cite{PhysRevLett.103.093902,NatPhys.6.192,NatPhys.10.394}

A central issue that frequently arises in the study of non-Hermitian
(and other) systems is the calculation of ground and excited states
together with their respective eigenenergies. Traditional methods such
as shooting, Newton and self-consistency schemes can be (in certain
cases) cumbersome to implement. In the presence of complex potentials some
methods are even ruled out as, e.g., imaginary time propagations
because the imaginary contributions add an oscillatory term to the
exponent and the algorithm does not converge. Reliable finite element
methods are possible but require initial guesses of high quality
\cite{IntJTheorPhys.54.4100}.  An effective and easy to implement
alternative is to use the so-called spectral renormalization method
\cite{JFluidMech.562.313,OptLett.30.2140,PhysicaD.184.276}, which
was successfully used on various problems including nonlocal
integrable and time-dependent systems
\cite{PhysRevLett.110.064105,StudApplMath.139.7,PhysicaD.358.15}.

The paper is organized as follows. In Sec.~\ref{OSR} we present a
detailed account of what we refer to as orthogonal spectral
renormalization method. In sections \ref{GS}, \ref{FES}, and
\ref{AES} the OSR method is explicitly constructed to compute the ground, first,
second, and generic excited states. Modifications and simplifications valuable
for practical applications are presented in Sec.~\ref{implementation}.
To demonstrate the applicability of the method we study some relevant
linear and nonlinear Schr\"odinger type operators, for which a high
number of states is calculated in Sec.~\ref{examples}. Conclusions are drawn
in Sec.~\ref{conclusion}.

\section{Orthogonal spectral renormalization}
\label{OSR} We begin our discussion by considering a general eigenvalue
problem in the form
\begin{equation}
  \label{eig} L\psi + f(\psi ) = E \psi\;,
\end{equation}
where $L$ is a linear differential operator and $f(\psi)$ is some nonlinear
function of the real or complex function $\psi.$ The problem considered here
is posed on the spatial domain $\Omega$ which is either bounded, the whole
real line, or the entire space for multi-dimensional problems.
Equation \eqref{eig} is supplemented with periodic or vanishing boundary
conditions. Furthermore, it is assumed that the eigenvalue problem
\eqref{eig} admits a ground state and $N$ excited states which we
respectively denote by $\psi_g, \psi_1, \cdots, \psi_N.$ Thus we have
\begin{equation}
  \label{eig-1} L\psi_j + f(\psi_j ) = E_j \psi_j\;,\;\;\;\;\; j=g,
  1,2,\cdots, N \;.
\end{equation}
Note that in the presence of a nonlinear term, there are of course 
many different notions of eigenvalue problems associated with Eq.\
\eqref{eig-1}. Since we are interested in quantum mechanical systems, it is in
addition assumed that all eigenfunctions $\psi_j$ are square integrable and
satisfy the normalization condition
\begin{equation}
  \label{normalization} \int_\Omega dx |\psi_j|^2 = 1 \;,\;\;\;\;\; j=g,
  1,2,\cdots, N\; .
\end{equation}
Importantly, it is further assumed that all eigenfunctions are mutually
orthogonal with respect to some type of inner product that accompany the
eigenvalue problem \eqref{eig}. Thus, this orthogonality condition is
abstractly given by
\begin{equation}
  \label{orthogonality-assumption} 
  \langle \psi_j, \psi_\ell \rangle  = 0 \;,\;\;\; j,\ell=g,
  1,2,\cdots,N; \;\;\; j\ne\ell \;.
\end{equation}
It should be pointed out that Sturm-Liouville theory for self-adjoint
(Hermitian) {\it linear} eigenvalue problems guarantees the existence of a
set of mutually orthogonal eigenstates. However, this is \textit{not}
the case for a linear non-Hermitian as well as for nonlinear (Hermitian or
not) boundary value problems. This explains the need for the orthogonality
condition \eqref{orthogonality-assumption}.

In this paper, our main interest is the numerical computation of the ground
and excited states $\psi_j$ and their respective eigenvalues (energies)
$E_j, j=g, 1,2,\cdots, N$. There are several well established numerical
methods that one can use to accomplish that goal. This includes
Newton's, shooting, marching and self-consistency methods to name a
few \cite{pang_computational_physics}. In 2005 the spectral renormalization
(SR) scheme was proposed to compute (ground) bound states for nonlinear systems
\cite{JFluidMech.562.313,OptLett.30.2140,PhysicaD.184.276}. However,
when one attempts to use this method to numerically find higher-order
modes it usually fails. Here, we propose an alternative SR scheme that
enables one to compute excited states and their respective
eigenvalues. The core idea is to implement the traditional SR
algorithm interfaced with a Gram-Schmidt type orthogonalization
procedure that prevents the scheme from converging to an undesired
mode. We term this method orthogonal spectral renormalization. In what
follows we outline the major steps in implementing this idea.

We introduce an unknown sequence of renormalization parameters $r_j$
(different from zero) and their respective renormalized wave functions
$\varphi_j$ via the change of variables
\begin{equation}
  \label{sub} \psi_j (x) = r_j \varphi_j (x)\;,\;\;\;\;\; j=g,
  1,2,\cdots, N\;.
\end{equation}
From \eqref{normalization} it follows that the renormalization factors $r_j$
satisfy the relation
\begin{equation}
  \label{normalization-1} |r_j|^2 = \frac{1}{\int_\Omega dx |\varphi_j
    (x)|^2} \;,\;\;\;\;\; j=g, 1,2,\cdots, N\;.
\end{equation}
The renormalized wave functions $\varphi_j$ satisfy the following boundary
value problem induced from Eq.~\eqref{eig}
\begin{equation}
  \label{eig-new} L\varphi_j + \frac{1}{r_j} f(r_j\varphi_j ) = E_j
  \varphi_j \;,\;\;\;\;\; j=g, 1,2,\cdots, N\;.
\end{equation}
With this at hand we next turn our focus to the question of how to devise an
algorithm to approximate the renormalized eigenfunctions $\varphi_j$, their
corresponding eigenvalues $E_j$ and normalizations $r_j$.
We shall denote by $\langle u,v
\rangle$ the inner product defined for any two complex-valued square integrable
functions $u$ and $v$ defined by
\begin{equation}
  \label{inner-product} \langle u,v \rangle = \int_\Omega u(x)
  \bar{v}(x) dx\;.
\end{equation}
Definition \eqref{inner-product} is usually adopted when dealing with
  self-adjoint eigenvalue problems. As we shall see in Sec.\
  \ref{non-her-inner-prod}, a different type of inner product is used for
  non-Hermitian systems. As mentioned above, bar denotes complex conjugation.
The induced norm is given by $||u||^2=\langle u,u \rangle$. Taking the inner
product of Eq.~\eqref{eig-new} with $\varphi_j$ gives an expression for the
eigenvalues $E_j$ for all $j=g, 1,2,\cdots, N$,
\begin{equation}
  \label{eig-E} E_j = \frac{\langle \varphi_j,L\varphi_j
    \rangle}{||\varphi_j||^2} + \frac{1}{r_j ||\varphi_j||^2} \langle
  \varphi_j,f(r_j\varphi_j ) \rangle \;.
\end{equation}
We remark that solutions to Eq.\ \eqref{eig} can be obtained using calculus of
variation. Indeed, the problem can be formulated as finding the ground state
that minimizes a suitable functional subject to the constraint $||\psi_g||=1.$
Similarly, the excited state is found by minimizing the same functional subject
to the constraint in the orthogonal complement of the lower states.

Next, we proceed with the task of computing the ground, first, and $N$th
excited state. The first step is to outline the computation of the ground state
-- which is necessary to obtain excited states.

\subsection{Ground state}
\label{GS}
The renormalized ground state $\varphi_g$ satisfies the following
eigenvalue problem which is obtained from Eq.~\eqref{eig-new}
\begin{equation}
  \label{eig-g} L\varphi_g + \frac{1}{r_g} f(r_g\varphi_g ) = E_g
  \varphi_g \;.
\end{equation}
The ground state is numerically found from the fixed point iteration
\begin{equation}
  \label{eig-g-iter} \varphi^{(n+1)}_g = \frac{1}{r^{(n)}_g}
  \left(L-E^{(n)}_g\right)^{-1} f\left(r^{(n)}_g\varphi^{(n)}_g \right)
  \;,
\end{equation}
where $n=1,2, \cdots$. The ground state eigenvalue $E_g$ is given by
\begin{equation}
  \label{eig-E-ground} E^{(n)}_g = \frac{\langle
    \varphi^{(n)}_g,L\varphi^{(n)}_g \rangle }{||\varphi^{(n)}_g||^2} +
  \frac{1}{r^{(n)}_g ||\varphi^{(n)}_g||^2} \langle
  \varphi^{(n)}_g,f(r^{(n)}_j\varphi^{(n)}_g ) \rangle \;,
\end{equation}
with
\begin{equation}
  \label{r-ground} |r^{(n)}_g|^2 \equiv \frac{1}{||\varphi^{(n)}_g||^2}
  \;.
\end{equation}
Upon convergence, the ground state solution for Eq.~\eqref{eig-1} is given by
\begin{equation}
  \label{ground-sol} \psi_g = r^{(\infty)}_g \varphi^{(\infty)}_g \;.
\end{equation}
Next, we explain how to use this information to calculate the first excited
state.

\subsection{First excited state}
\label{FES}
The first renormalized excited state $\varphi_1$ satisfies the boundary value
problem
\begin{equation}
  \label{eig-1-excite} L\varphi_1 + \frac{1}{r_1} f(r_1\varphi_1 ) = E_1
  \varphi_1 \;.
\end{equation}
If one implements the algorithm outlined in Sec.~\ref{GS}, the result of the
iterative process would be the ground state. To force the iteration to
``go away'' from the ground state we introduce a new renormalized excited
state $\eta_1$ defined by
\begin{equation}
  \label{eta-1} \varphi_1 = \eta_1 - c_g \psi_g \;,
\end{equation}
where the ``constant'' $c_g$ is given by
\begin{equation}
  \label{cg} c_g \equiv \frac{\langle \psi_g, \eta_1 \rangle
  }{||\psi_g||^2} \;.
\end{equation}
Notice that this choice of the parameter $c_g$ would force the first excited
state to be orthogonal (with respect to the inner product given in
\eqref{inner-product}) to the ground state. Indeed, we have
\begin{equation}
  \label{orth} \langle \psi_g, \varphi_1 \rangle =0\;.
\end{equation}
Since $\psi_1 =r_1\varphi_1,$ it follows that the ground and first excited
states of the system are orthogonal as well. The auxiliary function $\eta_1$
satisfies the eigenvalue problem
\begin{equation}
  \label{eig-1-excited} L \eta_1 + \frac{1}{r_1} f(r_1 \varphi_1)
  -E_1\eta_1 = c_g (L - E_1) \psi_g \;.
\end{equation}
As a functional of the new renormalized wave function $\eta_1,$ the first
excited state eigenvalue is given by (see Eq.~\eqref{eig-1-excite})
\begin{equation}
  \label{first-eig-E} E_1 = \frac{\langle
    \varphi_1,L\varphi_1\rangle }{||\varphi_1||^2} + \frac{1}{r_1
    ||\varphi_1||^2} \langle \varphi_1,f(r_1\varphi_1 ) \rangle \;,
\end{equation}
where
\begin{equation}
  \label{normalization-1-1} |r_1|^2 = \frac{1}{\int_\Omega dx | \eta_1 -
    c_g \psi_g |^2}\;.
\end{equation}

Equations \eqref{eig-1-excited} -- \eqref{normalization-1-1} are then
numerically solved with the aid of the following fixed point iteration
\begin{equation}
  \label{cg-it} c^{(n)}_g \equiv \frac{\langle \psi_g, \eta^{(n)}_1
    \rangle }{||\psi_g||^2} \;,
\end{equation}
\begin{equation}
  \label{eta-1-it} \varphi^{(n)}_1 = \eta^{(n)}_1 - c^{(n)}_g \psi_g \;,
\end{equation}
\begin{equation}
  \label{normalization-1-1-it} |r^{(n)}_1|^2 = \frac{1}{\int_\Omega dx |
    \eta^{(n)}_1 - c^{(n)}_g \psi_g |^2}\;,
\end{equation}
\begin{equation}
  \label{first-eig-E-it} E^{(n)}_1 = \frac{\langle
    \varphi^{(n)}_1,L\varphi^{(n)}_1 \rangle }{||\varphi^{(n)}_1||^2} +
  \frac{1}{r^{(n)}_1 ||\varphi^{(n)}_1||^2}
  \langle\varphi^{(n)}_1,f(r^{(n)}_1\varphi^{(n)}_1 )\rangle \;,
\end{equation}
\begin{equation}
  \label{eig-1-excited-it} \eta^{(n+1)}_1 = - \frac{1}{r^{(n)}_1} (L -
  E^{(n)}_1)^{-1} f(r_1 \varphi^{(n)}_1) + c^{(n)}_g \psi_g \;.
\end{equation}
Thus, the implementation of the orthogonal spectral renormalization
algorithm goes as follows: We first give an initial guess
$\eta^{(1)}_1(x)$ and compute the ``constant'' $c_g^{(1)}$ from
Eq.~\eqref{cg-it}. From Eq.~\eqref{eta-1-it} we have $\varphi^{(1)}_1$
which is then used in Eqs.~\eqref{normalization-1-1-it} and
\eqref{first-eig-E-it} to obtain approximations for the renormalization
constant $r_1^{(1)}$ and eigenvalue $E_1^{(1)}.$ The renormalized eigenfunction
$\eta_1^{(1)}$ is then updated using Eq.~\eqref{eig-1-excited-it}.

\subsection{$N$th excited state}
\label{AES}
The computation of an arbitrary excited state can be constructed from
knowledge of the ground and previous higher-order modes. By denoting
$\varphi_N,$ the $N^{{\rm th}}$ renormalized excited state, we have
\begin{equation}
  \label{eig-N-excite} L\varphi_N + \frac{1}{r_N} f(r_N\varphi_N ) = E_N
  \varphi_N \;.
\end{equation}
Following the Gram-Schmidt orthogonalization process, we define a new
renormalized excited state $\eta_N$ by
\begin{equation}
  \label{eta-N} \varphi_N = \eta_N - \sum_{j=g}^{N-1}c_j \psi_j \;,
\end{equation} with
\begin{equation}
  \label{cgN} c_j \equiv \frac{\langle \psi_j, \eta_N\rangle
  }{||\psi_j||^2}\;, \;\;\;\; j=g,1,2, \cdots, N-1\;.
\end{equation}
As a result, we have the orthogonality condition
$\langle \varphi_N, \psi_\ell \rangle =0,\;\; \ell = g, 1,2, \cdots,
N-1.$ It can be shown that the renormalized function $\eta_N$
satisfies the following boundary value problem
\begin{equation}
  \label{eig-N-excited} L \eta_N + \frac{1}{r_N} f(r_N \varphi_N) -
  E_N\eta_N = \sum_{\ell =g}^{N-1}c_\ell (L - E_N) \psi_\ell\;.
\end{equation}
The eigenvalue corresponding to the $N^{{\rm th}}$ excited state is given by
\begin{equation}
  \label{N-eig-E} E_N = \frac{\langle \varphi_N,L\varphi_N \rangle
  }{||\varphi_N||^2} + \frac{1}{r_N ||\varphi_N||^2} \langle
  \varphi_N,f(r_N\varphi_N ) \rangle \;,
\end{equation} where
\begin{equation}
  \label{normalization-N} |r_N|^2 = \frac{1}{\int_\Omega dx | \eta_N -
    \sum_{j=g}^{N-1}c_j \psi_j |^2}\;.
\end{equation}

To obtain a numerical approximation for the $N^{{\rm th}}$ excited
state we iterate the following system of equations until convergence
is achieved (here, $n=1,2,3, \cdots $)
\begin{equation}
  \label{cgN-it} c^{(n)}_j \equiv \frac{\langle \psi_j,
    \eta^{(n)}_N\rangle }{||\psi_j||^2}\;, \;\;\;\; j=g,1,2, \cdots,
  N-1\;,
\end{equation}
\begin{equation}
  \label{normalization-N-it} |r^{(n)}_N|^2 = \frac{1}{\int_\Omega dx |
    \eta^{(n)}_N - \sum_{j=g}^{N-1}c^{(n)}_j \psi_j |^2}\;,
\end{equation}
\begin{equation}
  \label{eta-N-it} \varphi^{(n)}_N = \eta^{(n)}_N -
  \sum_{j=g}^{N-1}c^{(n)}_j \psi_j \;,
\end{equation}
\begin{equation}
  \label{N-eig-E-it} E^{(n)}_N = \frac{\langle
    \varphi^{(n)}_N,L\varphi^{(n)}_N \rangle }{||\varphi^{(n)}_N||^2} +
  \frac{1}{r^{(n)}_N ||\varphi^{(n)}_N||^2} \langle
  \varphi^{(n)}_N,f(r^{(n)}_N\varphi^{(n)}_N ) \rangle \;,
\end{equation}
\begin{equation}
  \label{eig-N-excited-it} \eta^{(n+1)}_N = - \frac{1}{r^{(n)}_N} (L -
  E^{(n)}_N)^{-1} f(r^{(n)}_N \varphi^{(n)}_N) + \sum_{\ell
    =g}^{N-1}c^{(n)}_\ell \psi_\ell\;.
\end{equation}

\section{Numerical implementation}
\label{implementation} The basic three steps necessary for
implementing the OSR method are: (i) renormalization, (ii)
orthogonalization and (iii) fixed point iteration. The latter is
obtained by converting the underlying linear or nonlinear boundary
value problem to a fixed point (differential or integral)
equation. However, as is well known, there is no unique way to
reformulate a given eigenvalue problem into a fixed point equation. As
such, the preferred choice is dictated by the computational efficiency
and algorithm optimization.

\subsection{Possible simplifications for nonlinear problems}
As an example, for nonlinear problems with a Schr\"odinger type linear
part we have
\begin{equation}
  \label{L-op} L = -\frac{\partial^2}{\partial x^2} +V(x)\;,
\end{equation}
where $V(x)$ is either real or a complex valued potential. In this case the
OSR scheme is based on the nonlinear equation
\begin{multline} \left ( -\frac{\partial^2}{\partial x^2} +V(x)
  \right ) \eta_N + \frac{1}{r_N} f(r_N \varphi_N) - E_N\eta_N \\ =
  \sum_{\ell =g}^{N-1}c_\ell (L - E_N) \psi_\ell\;.
  \label{N-no-inversion}
\end{multline}

The iteration of this equation with the procedure presented in
Eq.~\eqref{eig-N-excited-it} can be costly (due to the required
inversion of the differential operator) and spectral methods can be
also cumbersome to implement. An alternative form to the fixed point
iteration \eqref{eig-N-excited-it} would be to take the Fourier
transform of Eq.~\eqref{N-no-inversion}, which results in the following new
fixed point equation
\begin{multline}
  \label{iteration-fourier} \hat{\eta}^{(n+1)}_N = \frac{(E^{(n)}_N
    +\xi^2)\hat{\varphi}^{(n)}_N} {k^2 +\xi^2} + \sum_{\ell
    =g}^{N-1}c^{(n)}_\ell \hat{\psi}_\ell \\ - \frac{1}{k^2 +\xi^2} F
  \left( V(x)\varphi^{(n)}_N + \frac{1}{r^{(n)}_N} f(r^{(n)}_N
    \varphi^{(n)}_N) \right) \; .
\end{multline}
The Fourier transform is denoted by $F$ and the
functions $\hat{\eta}$, $\hat{\varphi}$, and $\hat{\psi}$ are the
Fourier transforms of $\eta$, $\varphi$, and $\psi$, respectively. We
have rewritten $-\partial^2/\partial x^2 - E_N^{(n)}$ as
$-\partial^2/\partial x^2 +\xi^2 - E_N^{(n)} - \xi^2$ with $\xi^2$
being an arbitrary positive number such that $-\partial^2/\partial x^2
+\xi^2$ is positive definite.

In Eq.~\eqref{iteration-fourier} it can be clearly seen that the term
$\sum_{\ell =g}^{N-1}c^{(n)}_\ell \hat{\psi}_\ell$ only adds contributions to
$\eta$ which are orthogonal with respect to the inner product
\eqref{inner-product}. At the end we are not interested in these since we want
to determine $\varphi_N$. Thus, one can, in practical applications, do without
these terms and avoid the computation of their Fourier transforms. The
simplified iteration reads
\begin{subequations}
  \begin{multline}
    \label{iteration-fourier-simple} \hat{\Phi}^{(n+1)}_N =
    \frac{(E^{(n)}_N +\xi^2)\hat{\varphi}^{(n)}_N} {k^2 +\xi^2} \\ -
    \frac{1}{k^2 +\xi^2} F \left( V(x)\varphi^{(n)}_N +
      \frac{1}{r^{(n)}_N} f(r^{(n)}_N \varphi^{(n)}_N) \right) \; ,
  \end{multline}
  \begin{equation}
    \label{iteration-fourier-simple2} \hat{\varphi}^{(n+1)}_N =
    \hat{\Phi}^{(n)}_N - \sum_{\ell =g}^{N-1}c^{(n)}_\ell \hat{\psi}_\ell \; ,
  \end{equation}
\end{subequations}
and $\hat{\Phi}^{(n)}_N$ is used instead of $\hat{\eta}^{(n)}_N$ in all other
steps. Note that in this way we cannot obtain the auxiliary wave function
$\eta$. This will affect only the wave functions $\hat{\Phi}_N$, however,
the desired results $\varphi_N$ will be identical with those obtained from
Eq.~\eqref{iteration-fourier}.
For the numerical results reported in this paper, we used a fast
Fourier transform and the Fourier fixed point iteration
\eqref{iteration-fourier-simple} rather than inverting the operator
$-\partial^2/\partial x^2 +V(x) - E^{(n)}_N$.

\subsection{Renormalization for linear operators}
While the spectral renormalization was originally introduced for
nonlinear operators \cite{JFluidMech.562.313,OptLett.30.2140,
  PhysicaD.184.276} the iteration according to
Eq.~\eqref{iteration-fourier-simple} can also be applied to linear
operators, i.e., $f(r^{(n)}_N \varphi^{(n)}_N)=0$. In this case the
renormalization factor $r_N^{(n)}$, calculated from
Eq.~\eqref{normalization-N-it},  no longer appears in
Eq.~\eqref{iteration-fourier-simple} and it seems the renormalization
is no longer necessary. However, in practical applications it is
anyway necessary to renormalize such that the wave function will
neither grow above all limits nor converge to vanishing norm.

Thus, for a linear problem the procedure is outlined below. Again, we use a
simplified iteration based on Eqs. \eqref{iteration-fourier-simple} and
\eqref{iteration-fourier-simple2}, i.e.\ $\hat{\Phi}^{(n)}_N$ is used instead
of $\hat{\eta}^{(n)}_N$. First, the coefficients $c_j^{(N)}$ are computed
according to Eq.~\eqref{cgN-it}, then the normalization constant
$r_N^{(n)}$ is determined via Eq.~\eqref{normalization-N-it}, and
$\varphi_N^{(n)}$ is calculated from
Eq.~\eqref{iteration-fourier-simple2}. After this, it is the best
choice to introduce
\begin{equation}
  \psi_N^{(n)} = r_N^{(n)} \varphi_N^{(n)} \;, 
\end{equation}
and to use the modification
\begin{equation}
  E^{(n)}_N = \langle \psi^{(n)}_N,L \psi^{(n)}_N \rangle \;,
\end{equation}
of Eq.~\eqref{N-eig-E-it} to calculate the iterate of the energy eigenvalue
$E^{(n)}_N$. Finally the next step in the iteration is obtained from the
corresponding adaptation of Eq.~\eqref{iteration-fourier-simple}, viz.\
\begin{equation}
  \hat{\Phi}^{(n+1)}_N = \frac{(E^{(n)}_N +\xi^2)\hat{\psi}^{(n)}_N}
  {k^2 +\xi^2} - \frac{1}{k^2 +\xi^2} F \left ( V(x)\psi^{(n)}_N \right ).
\end{equation}

\subsection{Inner product for non-Hermitian potentials}
\label{non-her-inner-prod}
As a final comment, we presented the OSR scheme using the inner
product given in Eq.~\eqref{inner-product} which is suitable for a
self-adjoint linear operator $L$. For non-Hermitian systems, such as
the \PT-symmetric one discussed in this paper, a more suitable inner
product would be
\begin{equation}
  \label{inner-product-CPT} \langle u,v \rangle = \int_\Omega \left (
    \mathcal{CPT} u(x) \right ) v(x) dx \;,
\end{equation} where the $\mathcal{C}$ operator is defined in
Refs.~\cite{PhysRevLett.89.270401,ContempPhys.46.277}.

\section{Examples}
\label{examples}
In this section we apply the OSR method to various linear and nonlinear
eigenvalue problems including \PT-symmetric ones.

\subsection{Linear Hermitian systems}
\subsubsection{Harmonic oscillator}
\begin{figure}[t]
  \includegraphics[width=\columnwidth]{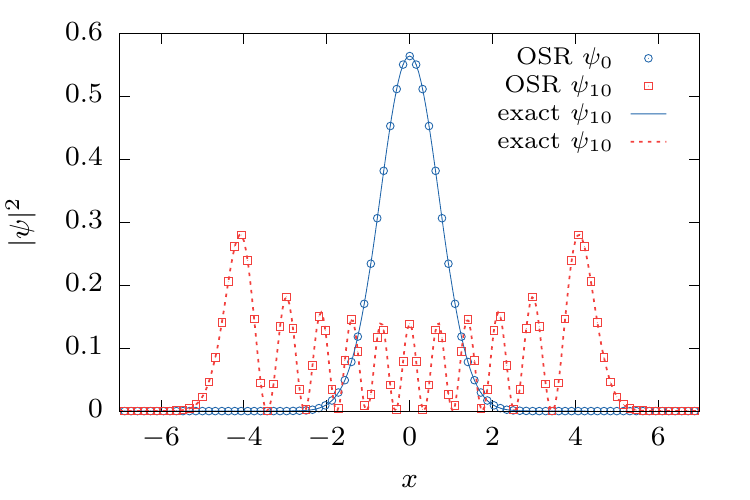}
  \caption{\label{fig:ho_1d}Eigenfunctions $\psi_0$ and $\psi_{10}$ of the
    1d harmonic oscillator \eqref{eq:ho_1d} calculated with the orthogonal
    spectral renormalization method in comparison with the exact solution.}
\end{figure}

The first example we consider is the one-dimensional simple quantum harmonic
oscillator. The dimensionless Schr\"odinger equation reads
\begin{equation}
  \left[-\frac{\partial^2}{\partial x^2} + x^2\right]\psi(x) = E\psi(x)\,.
  \label{eq:ho_1d}
\end{equation}
We choose a field of view of $x \in [-6,6]$ with a resolution of
$M=128$ points and a singularity prevention constant of $\xi^2 =
15$. The algorithm is quite insensitive about the initial guesses,
therefore one can always choose the initial function $\varphi_0 =
\e^{-x^2}$ for all states. With this set of parameters it is possible to
calculate the first 9 states up to machine precision (14 to 15 valid digits)
within $N_{\mathrm{iter}}\approx 230$--$1230$ iterations. To calculate
higher states up to machine precision, the resolution, the field of
view and the singularity prevention constant have to be increased. For
example a parameter set of $M=256,$ $x \in [-20,20]$ and $\xi^2 =
1000$ allows for a calculation of the first $100$ states up to machine
precision. In this case the iteration count is
$N_{\mathrm{iter}}\approx 11000$--$15000$.  Exemplarily the resulting
eigenfunctions $\varphi_0$ and $\varphi_{10}$ are shown in Fig.\
\ref{fig:ho_1d}
in comparison with the analytic solution.  It is remarkable that despite the
rather low resolution for the highly excited states the energy matches
perfectly the analytical result up to machine precision.

A calculation in two dimensions with
\begin{equation}
    \left[-\nabla^2 + x^2 + y^2\right]\psi(x,y) = E\psi(x,y) \;,
\end{equation}
shows similar results. The energy eigenvalues nicely converge to the
analytically known values. Excited states can be acquired up to machine
precision as well. However, if one wants to converge to the eigenstates
of the quantum numbers in the Cartesian basis, i.e., $n_x$ and $n_y$, and not
to some superpositions of these, even this can be achieved. In this case
one has to specify the initial guesses more precisely, e.g., for the first
states one can choose
\begin{align}
    \varphi_{00} &= \e^{-(x^2+y^2)}\,, &
    \varphi_{10} &= x\e^{-(x^2+y^2)}\,, \notag\\
    \varphi_{01} &= y\e^{-(x^2+y^2)}\,, &
    \varphi_{20} &= x^2\e^{-0.5(x^2+y^2)}\,, \notag\\
    \varphi_{11} &= xy\e^{-(x^2+y^2)}\,, &
    \varphi_{02} &= y^2\e^{-0.5(x^2+y^2)} \;,
\end{align}
to obtain the required nodal structure.

\subsubsection{Anharmonic oscillator}
\begin{figure}[t]
  \includegraphics[width=\columnwidth]{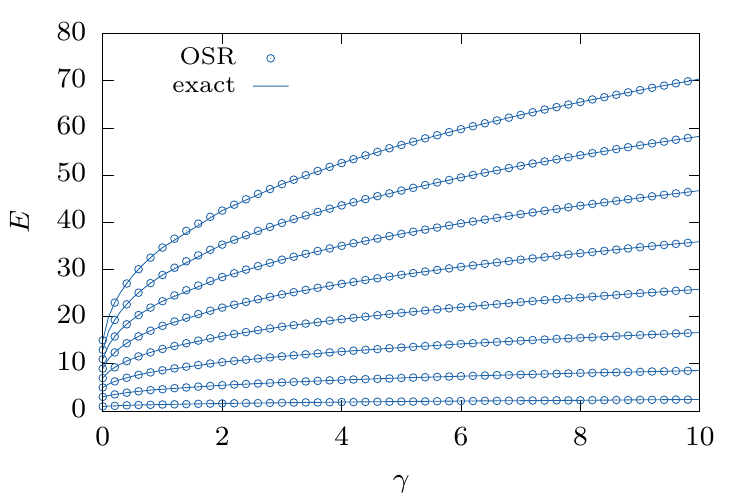}
  \caption{\label{fig:anharmonic}Spectrum of the first eight eigenstates of
    the anharmonic oscillator \eqref{eq:anharmonic} calculated with the
    orthogonal spectral renormalization in comparison with a numerically
    accurate calculation from \cite{JMathPhys.14.1190}.}
\end{figure}

The next example is an anharmonic quartic one-dimensional oscillator whose
Schr\"odinger equation reads
\begin{equation}
  \left[-\frac{\partial^2}{\partial x^2} + x^2 + \gamma x^4\right]\psi(x)
  = E\psi(x)\,.\label{eq:anharmonic}
\end{equation}
Here we choose $x \in [-8,8]$, $M=128,$ $\xi^2=5\cdot 10^{4}$ and
$\varphi_0 = \e^{-x^2}$. This allows for calculations up to machine
precision for the first ten states up to $\gamma = 10$. The spectrum
in dependence of $\gamma$ for the first eight states is shown in Fig.\
\ref{fig:anharmonic}
in comparison with exact calculations from \cite{JMathPhys.14.1190}. The
agreement is perfect.

\subsubsection{Particle in a box}
\begin{figure}[t]
  \includegraphics[width=\columnwidth]{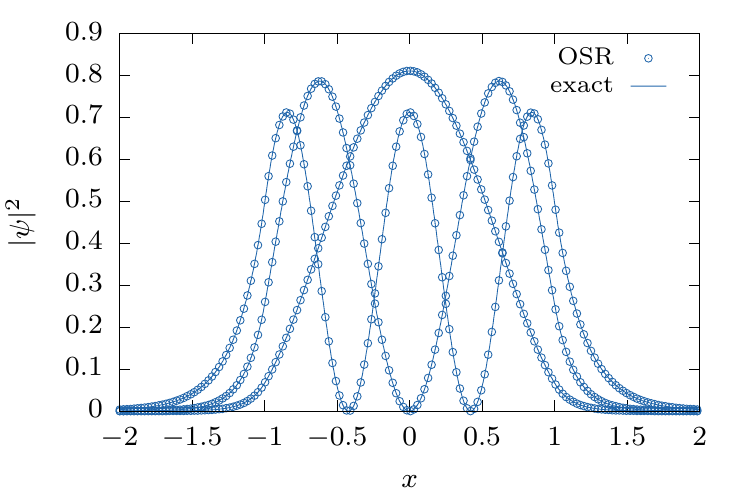}
  \caption{\label{fig:particle_in_box}Eigenfunctions of the particle in the
    box problem \eqref{eq:particle_in_box} with $a=1$ and $V_0=20$.
    The orthogonal spectral renormalization solution is compared with the
    analytical solution.}
\end{figure}

A third linear Hermitian example is a particle in a box with finite walls,
\begin{gather}
  \left[-\frac{\partial^2}{\partial x^2} + V(x)\right]\psi(x)
  = E\psi(x)\,, \notag \\
  V(x) = \begin{cases}
    0 & |x| < a \\
    V_0 & \text{otherwise}
  \end{cases}\,.
  \label{eq:particle_in_box}
\end{gather}
The analytical energy eigenvalues are determined by the transcendental
equations
\begin{align}
  \sqrt{\frac{V_0-E}{E}} &= \tan\left(\sqrt{E} a\right)\,,
  & \text{symmetric}\,,\\
  -\sqrt{\frac{E}{V_0-E}} &= \tan\left(\sqrt{E} a\right)\,,
  & \text{antisymmetric}\;.
\end{align}
We chose $a = 1$ and $V_0 = 20$, which leads to three bound states
with the energies
\begin{align}
    E_0 &= 1.63948\,, &
    E_1 &= 6.44188\,, &
    E_2 &= 13.8915\,.
\end{align}
For the spectral renormalization algorithm, we choose $x \in [-5,5]$,
$M=4096,$ $\xi^2 = 10$ and an initial guess of $\varphi_0 = \e^{-x^2}$. This
yields the energy eigenvalues
\begin{align}
    E_0 &= 1.64012\,, &
    E_1 &= 6.44434\,, &
    E_2 &= 13.89634,
\end{align}
within $N_{\mathrm{iter}}\approx 60-160$ iteration steps. The
corresponding wave function in comparison with the analytical solution
can be found in Fig.\ \ref{fig:particle_in_box}.
For this example, the solution cannot be retrieved up to machine precision. In
addition, the accuracy of the solution depends on the field of view and the
resolution in a nontrivial way, therefore, a fine adjustment has to be
performed to retrieve the most accurate solutions. This may result
from the discontinuity at $x = a$ and the fact that the finite
discretization can only approximate such a point.

\subsection{Linear \PT-symmetric systems}
In this section we will demonstrate the application of the orthogonal
spectral renormalization algorithm on a linear \PT-symmetric
system. We use the well-known toy model of Bender and Boettcher
\cite{PhysRevLett.80.5243}
\begin{equation}
  \left[-\frac{\partial^2}{\partial x^2} - (\im x)^\epsilon\right]\psi(x)
  = E\psi(x)\label{eq:modified_ho}\,.
\end{equation}
For the parameter $\epsilon > 2$ the system possess an unbroken \PT
symmetry with an entirely real spectrum. For values $\epsilon < 2$ and
decreasing $\epsilon$ always two energy levels merge into a pair of
complex conjugate eigenvalues. Below $\epsilon = 1$ no real
eigenvalues exist. The value of $\epsilon = 2$ is the special case of
the harmonic oscillator.

\begin{figure}[t]
    \includegraphics[width=\columnwidth]{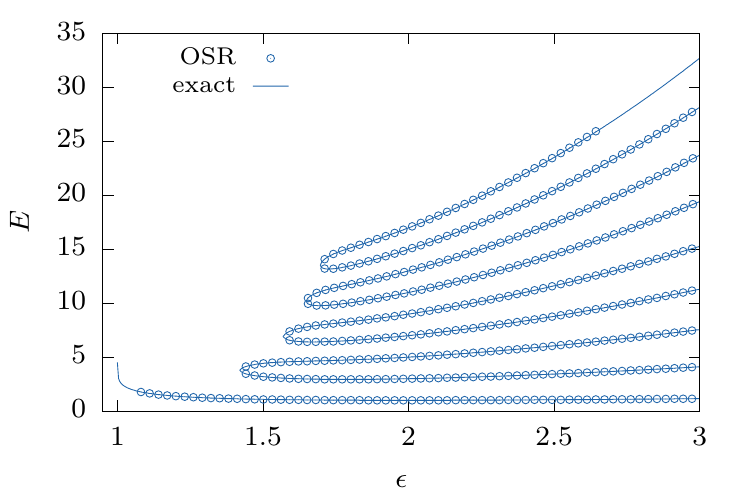}
    \caption{\label{fig:modified_ho}Spectrum of the first nine \PT-symmetric
      states of the complex extended harmonic oscillator model
      \eqref{eq:modified_ho}. The solution of the orthogonal spectral
      renormalization is compared with a numerically accurate solution. The
      algorithm fails to converge for the ground state for small $\epsilon$
      close to 1 and for higher excited states for larger $\epsilon$.}
\end{figure}

We choose the following set of parameters: $x \in [-8,8]$, $M = 128,$ 
$\xi^2=2\cdot 10^4$ and $\varphi_0 = \e^{-x^2}$. The resulting
spectrum for the \PT-symmetric solutions can be found in Fig.\
\ref{fig:modified_ho} compared with a numerical correct solution
calculated with a finite difference scheme. In the calculation for the
higher excited states, the \CPT product is used to project out the
states with lower chemical potential. The algorithm works quite well
to obtain the excited states. For increasing $\epsilon$ and higher
states the parameter $\xi^2$ has to be increased as well. This
explains the chosen high value of $\xi^2$ to converge for all selected
states.

However, there is a restriction that the value $\epsilon$ cannot be
increased much further than $\epsilon=3$. In addition for higher
excited states the algorithm fails to converge as well, even when the
renormalization factor $\xi^2$ is increased. This can be seen in the
Fig.\ \ref{fig:modified_ho}.
For higher excited states data points are missing. The algorithm also fails to
converge for the ground state for values of $\epsilon$ close to 1, in that
range where the accurate ground state energy begins to diverge.

\subsection{Nonlinear Hermitian system}
We now turn to nonlinear systems and study as an example the Gross-Pitaevskii
equation of a Bose-Einstein condensate in a harmonic trap, viz.\
\begin{equation}
  \left[-\frac{\partial^2}{\partial x^2} + x^2 + g|\psi|^2\right]\psi
  = E \psi \; ,
  \label{eq:GPE_harmonic}
\end{equation}
where $g$ measures the strength of the nonlinearity and $E$ is the
energy eigenvalue, which in the nonlinear problem has the physical meaning of
a chemical potential.

\begin{figure}[t]
  \includegraphics[width=\columnwidth]{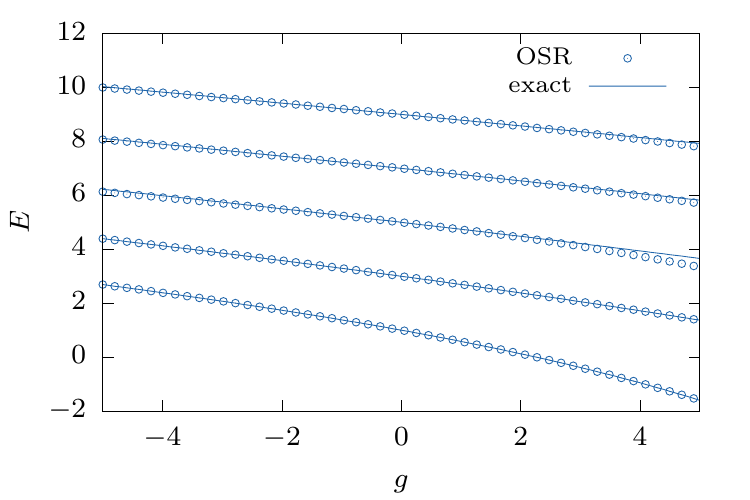}
  \caption{\label{fig:nonlin_ho_mu} Spectrum of a Bose-Einstein condensate
    in a harmonic trap in dependence of the nonlinearity parameter $g$.
    The chemical potentials $E$ are calculated with the orthogonal spectral
    renormalization method and compared with a numerically exact calculation
    in an oscillator basis.}
\end{figure}

\begin{figure*}[t]
  \includegraphics[width=\textwidth]{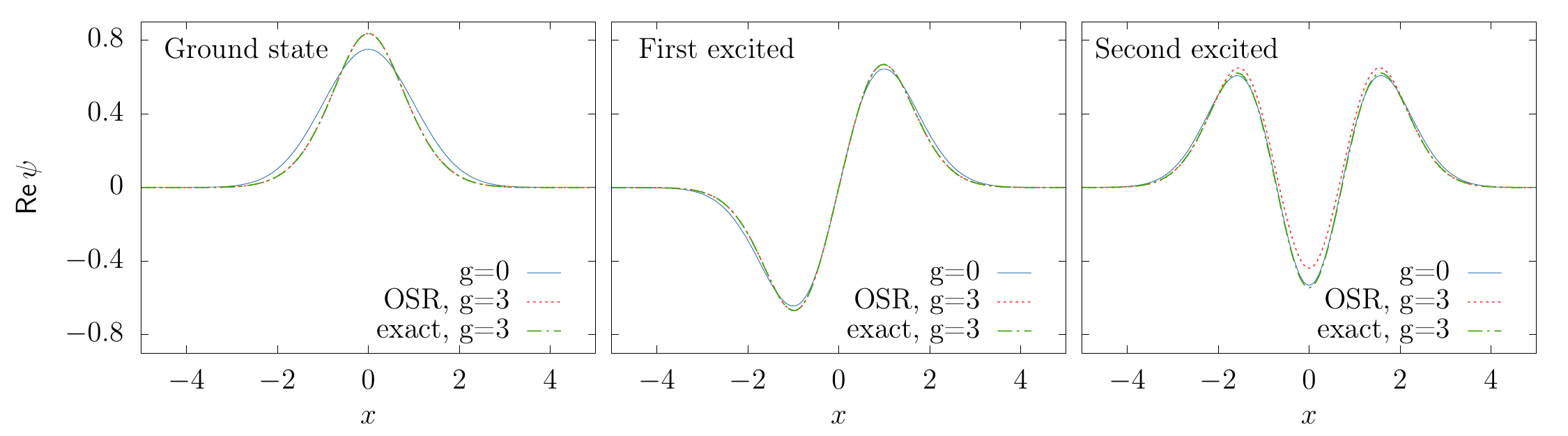}
  \caption{\label{fig:nonlin_waves} Wave functions of the Gross-Pitaevskii
    equation \eqref{eq:GPE_harmonic} for the ground state and the first two
    excited states calculated with the OSR and in an harmonic oscillator
    basis (exact).}
\end{figure*}

The chemical potentials for the ground state and the first four
excited states calculated with the orthogonal spectral renormalization are
shown in Fig.\ \ref{fig:nonlin_ho_mu}
in comparison with a calculation in an oscillator basis, of which we checked
that the values are numerically exact. The corresponding wave functions are
shown in Fig.\ \ref{fig:nonlin_waves}
and help to exemplify how the OSR works in such a nonlinear system. The initial
wave function of the OSR iteration was a simple Gaussian $\varphi_0 =
\e^{-x^2}$ and we used $x \in [-7,7]$, $M = 256,$ and $\xi^2=30$.

We can clearly observe that the ground state is found accurately.
The chemical potentials of both methods match perfectly. This is also
true for the first excited state. This result can be expected. As can be seen
in Fig.\ \ref{fig:nonlin_waves} the first excited state is antisymmetric with
respect to a reflection about $x=0$ whereas the ground state is symmetric.
Consequently the two states are orthogonal and the calculation of the first
excited state as outlined in Sec.\ \ref{FES} converges nicely to the correct
wave function.

For the second excited state the situation changes. The wave function is again
symmetric. In the nonlinear equation there is no need for it to be orthogonal
to the ground state, and in fact it is not. However, the orthogonalization
scheme of the OSR will force the wave function to be orthogonal to the
ground and first excited state. As a result the wave function does not converge
to the desired result. Anyway for small nonlinearities the result is still
a reasonable approximation, which loses in quality with increasing $g$. Since
the same reason exists for all higher excited states, for all of their chemical
potentials small deviations from the converged oscillator basis calculation
appear for large values of $g$.

\subsection{Nonlinear non-Hermitian systems}
\subsubsection{BEC in a double well with gain and loss}
The next example is a one-dimensional Bose-Einstein condensate in a
double-well potential with an additional complex gain-loss term. This system
was introduced and discussed in \cite{FortschrPhys.61.124} and is described
by the stationary Gross-Pitaevskii equation
\begin{equation}
  \left[-\frac{\partial^2}{\partial x^2} + V + g|\psi|^2\right]\psi
  = E \psi \; , 
\end{equation}
with a complex \PT-symmetric potential given by
\begin{equation}
  V = \omega^2 x^2 + V_0\e^{-\sigma x^2} + \im\Gamma x\e^{-\rho x^2}\,.
  \label{eq:double_well}
\end{equation}
The parameters used in our numerical study are
\begin{align}
    \rho &= \frac{\sigma}{2\ln\left(\frac{V_0\sigma}{\omega^2}\right)}\,,&
    \omega &= 0.5\,,&
    V_0 &= 4\,,&
    \sigma &= 0.5\,.
\end{align}
The real part of the potential represents the double-well trap and the
imaginary part a coherent in- and outcoupling of particles in the
individual wells. Here we choose the parameters $x \in [-7.5,7.5]$, 
$M= 128$ and $\xi^2=30$. The system has a \PT-symmetric ground state and
excited state as well as two complex conjugate \PT-broken states.

\begin{figure*}[t]
    \includegraphics[width=\textwidth]{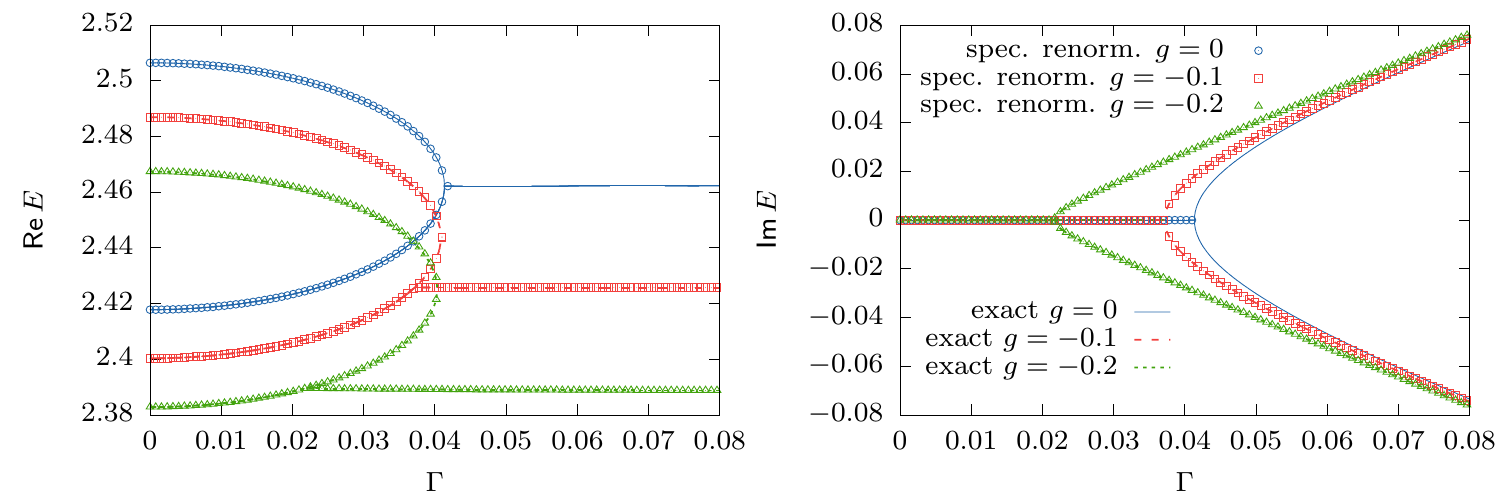}
    \caption{\label{fig:double_well}Spectrum of the double well
      \eqref{eq:double_well} calculated with the orthogonal spectral
      renormalization method in comparison with a solution from
      \cite{FortschrPhys.61.124}.}
\end{figure*}

The calculated spectrum as a function of the gain-loss parameter
$\Gamma$ can be found in Fig.\ \ref{fig:double_well}
compared with the numerically accurate solution from
\cite{FortschrPhys.61.124}. In the linear case for $g=0$ the ground and excited
states can be calculated perfectly. To calculate the excited state, the ground
state is subtracted via the \CPT product in each step. We use again a simple
Gaussian $\varphi_0 = \e^{-x^2}$ as initial guess. In the linear case
the system becomes \PT-broken at a value of $\Gamma \approx 0.04$. For
greater values there exist two complex conjugate \PT-broken
states. The spectral renormalization method fails in this area to
converge, independently of the initially chosen wave functions.

This changes if a nonlinearity $g\neq 0$ is introduced. Here the
\PT-broken states can be calculated if one chooses a suitable initial
\PT-broken guess, e.g.,
\begin{align}
    \varphi_+ = 1.1\e^{-(x-2)^2} + \e^{(x+2)^2}\,, \notag\\
    \varphi_- = \e^{-(x-2)^2} + 1.1\e^{(x+2)^2}\,.
\end{align}
In the nonlinear system the problem arises that there is a range in
which the \PT-broken states lie energetically below the \PT-symmetric
ground state. In this range one can enforce a convergence into a
\PT-symmetric state by dropping the imaginary part of the Fourier
transformed wave function in each iteration step.

Due to the nonlinearity the states are no longer orthogonal with
respect to the $\mathcal{CPT}$ product. Nevertheless, as the
nonlinearity is quite small, i.e. $|g| \leq 0.2$, the states are
approximately orthogonal, therefore, the acquired energy eigenvalues
are correct up to the third decimal digit.

\subsubsection{2d BEC excitations with gain and loss}
In a last example we consider a two-dimensional Bose-Einstein
condensate trapped in a harmonic trap with optional gain-loss
terms. We calculate ground and several excited states. The
Gross-Pitaevskii equation reads
\begin{equation}
  \left[-\nabla^2 + x^2 + y^2 + \im \Gamma x \e^{-(x^2+y^2)}
  + g|\psi|^2\right]\psi = E \psi\,.\label{eq:2d_bec}
\end{equation}
This represents a harmonic trap and a \PT-symmetric gain-loss term,
where $\Gamma$ is the strength of the in- and out coupling.
We choose $x \in [-5,5]$, $M=128$ and $\xi^2 = 30$. Initial guesses
are the harmonic oscillator ground and first excited states oriented
in $x$- and $y$-direction as well as a vortex ansatz:
\begin{align}
    \varphi_{\mathrm G} &= \e^{-(x^2+y^2)}\,, &
    \varphi_{\mathrm V} &= (x+\im y)\e^{-(x^2 + y^2)}\,,\notag\\
    \varphi_x &= x\e^{-(x^2+y^2)}\,, &
    \varphi_y &= y\e^{-(x^2+y^2)}\,.
\end{align}
Excited states are again calculated with the use of the \CPT product.

\begin{figure*}[t]
  \includegraphics[width=\textwidth]{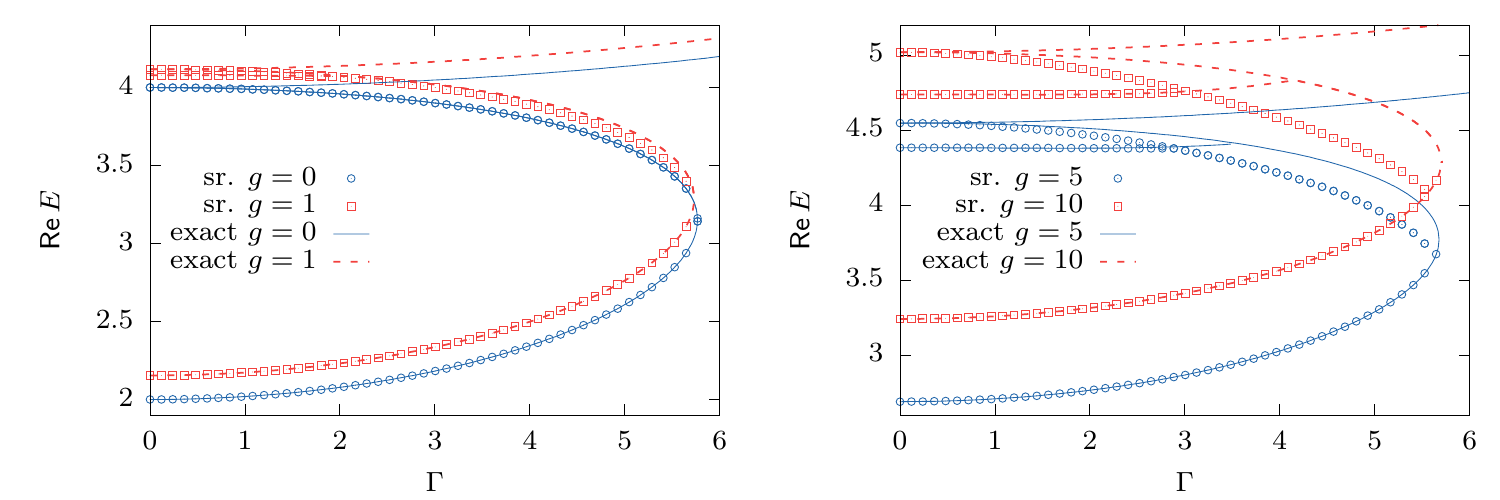}
  \caption{\label{fig:2d_bec}Spectrum of the lowest states of the two
    dimensional Bose-Einstein condensate in a harmonic trap with gain and
    loss \eqref{eq:2d_bec}. The orthogonal spectral renormalization solution
    is compared with a calculation in a harmonic oscillator basis
    \cite{PhysRevA.95.053613}.}
\end{figure*}

The resulting spectrum as a function of the gain-loss parameter
$\Gamma$ can be found in Fig.~\ref{fig:2d_bec}
in comparison with a numerically accurate solution calculated in a harmonic
oscillator basis \cite{PhysRevA.95.053613}. In the linear case $g=0$ the states
are perfectly orthogonal and the results match the accurate
solution. In the nonlinear case with $\Gamma = 0$ the states are
orthogonal as well and the result is again correct. In the nonlinear
non-Hermitian case the states are in general no longer orthogonal,
therefore, the spectral renormalization solutions deviate the more the
larger the nonlinearity $g$ is chosen. However, the state oriented in
$y$-direction is orthogonal to the ground state despite the
nonlinearity and non-Hermiticity, and this state can be calculated
correctly, independently of the chosen value of $g$ and $\Gamma$.

\section{Conclusion}
\label{conclusion}
In this paper, we developed and presented the orthogonal
spectral renormalization (OSR) method as a numerical scheme to compute
ground and excited states as well as their corresponding eigenvalues,
in all cases in which the states are orthogonal to each other
according to a certain inner product. The scheme can be used in both
linear and nonlinear boundary value problems. The OSR algorithm can be
described using the following major steps: (i) rewriting the given
eigenvalue problem in terms of a fixed point equation, (ii) introduce
a spectral renormalization parameter whose sole purpose is to control
the converge properties of the iteration, (iii) perform a Gram-Schmidt
orthogonalization process that forbids the iteration from converging
to an undesired (linear or nonlinear) mode; and (iv) compute the
solution sought using a fixed-point iterative scheme. The proposed
method extends the ``classical'' spectral renormalization scheme first
introduced in 2005 (to compute ground state) to enable the numerical
calculation of arbitrary excited states. The OSR has several
advantages: (i) it allows the flexibility to choose large varieties of
initial guesses without diverging, (ii) easy to code especially at
higher dimensions and (iii) it can easily handle problems with complex
and random potentials. 
The OSR method is implemented on typical {\it Hermitian} linear and
nonlinear eigenvalue problems where it proved to work very fast and reliably.
Examples include the linear harmonic and anharmonic oscillators,
one-dimensional particle in a box and the nonlinear
Schr\"odinger/Gross-Pitaevskii equation in the presence of a Hermitian
harmonic trap. In addition, the OSR scheme is used to compute the spectrum
of the \PT symmetric Hamiltonian from the seminal work of Bender and
Boettcher and a BEC in a double-well potential with gain and loss.

\end{document}